\documentclass[modern]{aastex}

\usepackage{natbib}
\usepackage{amsmath}
\usepackage{graphicx}
\usepackage{rotating}
\usepackage{epsfig}
\usepackage{epstopdf}
\usepackage{enumitem}

\newcommand{\Rs}{$ R_{\odot}$}

\newcommand{\Bk}{$B_{k}$}

\newcommand{\vcm}{cm$^{-3}$}
\newcommand{\paperi}{Paper \Rmnum{1}}

\makeatletter

\newcommand{\Rmnum}[1]{\expandafter\@slowromancap\romannumeral #1@}
\newcommand{\matr}[1]{\mathbf{#1}} 
\makeatother

\begin{document}

\shorttitle{An atlas of coronal electron density \Rmnum{2}}
\shortauthors{Morgan}
\title{An atlas of coronal electron density at 5\Rs\ \Rmnum{2}:\\
A spherical harmonic method for density reconstruction}
\author{Huw Morgan} 
\affil{Solar System Physics Group, Department of Physics, Aberystwyth University, Ceredigion, Cymru, SY23 3BZ}
\email{hmorgan@aber.ac.uk}

\begin{abstract}
This is the second of a series of three papers that present a methodology with the aim of creating a set of maps of the coronal density over a period of many years. This paper describes a method for reconstructing the coronal electron density based on spherical harmonics. By assuming a radial structure to the corona at the height of interest, line-of-sight integrations can be made individually on each harmonic basis prior to determining coefficients, i.e. the computationally-expensive integrations are calculated only once during initialization. This approach reduces the problem to finding the set of coefficients which best match the observed brightness using a regularized least-squares approach, and is very efficient. The method is demonstrated on synthetic data created from both a simple and an intricate coronal density model. The quality of reconstruction is found to be reasonable in the presence of noise and large gaps in the data. The method is applied to both LASCO C2 and STEREO COR2 coronagraph observations from 2009/03/20, and the results from both spacecraft compared. Future work will apply the method to large datasets.
\end{abstract}

\maketitle

\section{Introduction}

Reliable maps of the coronal density are important for linking various solar wind structures to the low solar atmosphere, for studies of the coronal response to the solar cycle, and for space weather applications: either as an inner boundary conditions for solar wind models, or for direct ballistic extrapolation into interplanetary space. Estimates of the coronal electron density can be made through inversion of coronal visible light observations. This has been achieved using several methods of varying complexity during eclipses, or by coronagraphs, for several decades. The introduction of \citet{morgan2015} gives a summary of the field, including discussion of the difficulties involved and examples of applications.  A comprehensive review is given by \citet{aschwanden2011b}.

This paper presents a new inversion method based on spherical harmonics for the extended inner solar corona, valid for regions where the large-scale structure is close to radial. Spherical harmonics as a basis for 3D reconstruction is used in some branches of medicine and geophysics \citep[e.g.][and references within]{merrill1996,arridge2009,levis2015}. The method is described in section \ref{method}, and is tested on a simple set of synthetic data in section \ref{simple}. A more complicated set of synthetic data is discussed in section \ref{complex}. An approach to regularizing the higher-order spherical harmonics is presented in section \ref{regular}. A discussion of datagaps, noise and temporal changes is given in section \ref{noise}. Application to observations are demonstrated in section \ref{observations}. Conclusions are in section \ref{conclusions}. The appendix presents an alternative method to calculate the spherical harmonic coefficients based on iteration rather than least-squares.

\section{Inversion using spherical harmonics}
\label{method}

\subsection{Outline}
\label{outline}
For a spherical surface at a constant height $r=r_0$, the coronal density $\rho$ at Carrington longitude $\phi$ and latitude $\theta$ may be approximated by a spherical harmonic basis,

\begin{equation}
\label{eq1}
\rho (\phi,\theta,r_0) = \sum_{i=0}^{n_{sph}-1} c_i S_i (\phi,\theta)
\end{equation}
where the $c_i$ are coefficients and $S_i$ are the real-valued spherical harmonics, with the $i$ index related to latitudinal order $l$ ($l \leq L$, where $L$ is the highest order) and longitudinal order $m$ ($-l \leq m \leq l$) by:
\begin{table}[h]
\centering
\begin{tabular}{ccc}
$i$ & $l$ & $m$ \\
\tableline
\\
0 & 0 & 0\\
1 & 1 & -1 \\
2 & 1 & 0 \\
3 & 1 & 1 \\
. & . & . \\
$n_{sph}-1$ & $L$ & $L$  \\
\end{tabular}
\end{table}

Note that $S_0$ is the mean density component (a constant at all $\phi$ and $\theta$) and $n_{sph}=(L+1)^2$. By increasing the order $L$ to large values, any sufficiently continuous density structure can be well approximated by equation \ref{eq1}. 

If a radial coronal density structure is assumed above the height of interest, the profile $f(r \geq r_0)$ of density with height can be described by a simple function. For example, considering mass flux conservation for a spherically-expanding corona under acceleration for heights at around 5\Rs, 
\begin{equation}
\label{eq2}
f(r)=\left( \frac{r_0}{r} \right)^\alpha, \hspace{3em} r \geq r_0
\end{equation}
with $\alpha=2.2$. Thus the coronal density can be described by 
\begin{equation}
\label{eq3}
\rho (\phi,\theta,r) = \rho (\phi,\theta,r_0) f(r), \hspace{3em} r \geq r_0
\end{equation}

For a volume segmented into discrete voxels, the observed K-coronal (electron) brightness \Bk\ is the line-of-sight summation of the product of density and a factor $g$ which contains known constants, Thomson scattering coefficients and the length of each line-of-sight segment through each voxel (see for example section 2.1 of \citet{quemerais2002}, and references within):
\begin{equation}
\label{eq5}
B_k = \sum_{j=1}^{n_{los}}g_j \rho_j = \sum_{j=1}^{n_{los}}g_j f(r_j)\sum_{i=0}^{n_{sph}} c_i S_{ij},
\end{equation}
where the $j$ index labels voxels lying along the line of sight, thus $S_{ij}$ is the value of the spherical harmonic at order level $i$ and voxel $j$.

Each spherical harmonic $S_{ij}$ may be summed independently of the other harmonics along the line of sight to give the brightness contribution resulting from each harmonic. Defining $A_i$:
\begin{equation}
\label{eq6}
A_i = \sum_{j=1}^{n_{los}}g_j f(r_j) S_{ij},
\end{equation}
the total brightness is given by
\begin{equation}
\label{eq7}
B_k = \sum_{i=0}^{n_{sph}} c_i  A_i.
\end{equation}

This describes a linear relationship between the contribution from each spherical harmonic density distribution and the observed brightness. For the purpose of finding an unknown density distribution from observed brightness, a reconstruction space with prescribed $S_{ij}$, $f(r_{j})$ and $g_j$ is created. The line of sight summations of equation \ref{eq6} are calculated, and the problem is reduced to finding the coefficients $c_i$ - thus the line-of-sight integrations are made only once, leading to high efficiency. Given a large number of observations ($n_{obs} \gg n_{sph}$), the system is overdetermined and can be solved using least squares. The ability to perform the line of sight summations independently for each spherical harmonic is based on the assumption of a radially-structured corona at heights above the height of interest, and a uniform profile to the decrease in density with height (e.g. equation \ref{eq2}). The assumption of a radial corona is reasonable at $r=$5\Rs, and the approximation of an assumed radial density profile is discussed later.

\subsection{Application}
Consider a set of observed coronal images recording brightness $B_k$, taken over an extended time period (e.g. half a solar rotation, $\sim$2 weeks). Circular samples of data at constant distance from Sun center, at a height at which the coronal structure is deemed close to radial (e.g. 5\Rs), are extracted over the time period, giving $b$, which records $B_k$ as a function of position angle and time.  For each member of $b$, a geometrical line-of-sight is defined through the corona, extending to large heights behind and in front of the point of closest approach to the Sun (similar to the description in the following section for the creation of synthetic observations). A set of $S_{ij}$, $g_j$, and $f(r)$ are prepared (with the unknown $f(r)$ set according to equation \ref{eq2}). The line-of-sight summation of equation \ref{eq6} is then implemented. This gives a set $A_{i}$, one for each spherical harmonic, each of size $n_{obs}$. Assuming a normal distribution to observational errors, the problem is reduced to solving
\begin{equation}
\label{eq8}
\min_c \; |\matr{b-Ac}|^2,
\end{equation}
with matrix $\matr{A}$ of size $n_{sph} \times n_{obs}$, $\matr{b}$ of size $n_{obs}$ and $\matr{c}$ the coefficients of size $n_{sph}$. The least-squares solution to equation \ref{eq8} is
\begin{equation}
\label{eq10}
\matr{c=(A^\intercal A)^{-1}A^\intercal b}.
\end{equation}
For numerical stability, before applying equation \ref{eq10}, $\matr A$ and $\matr b$ are divided by the mean of the absolute values of $\matr A$ (both contain very small numbers).

\section{A simple test}
\label{simple}

\begin{figure}[h]
\begin{center}
\includegraphics[width=8.5cm]{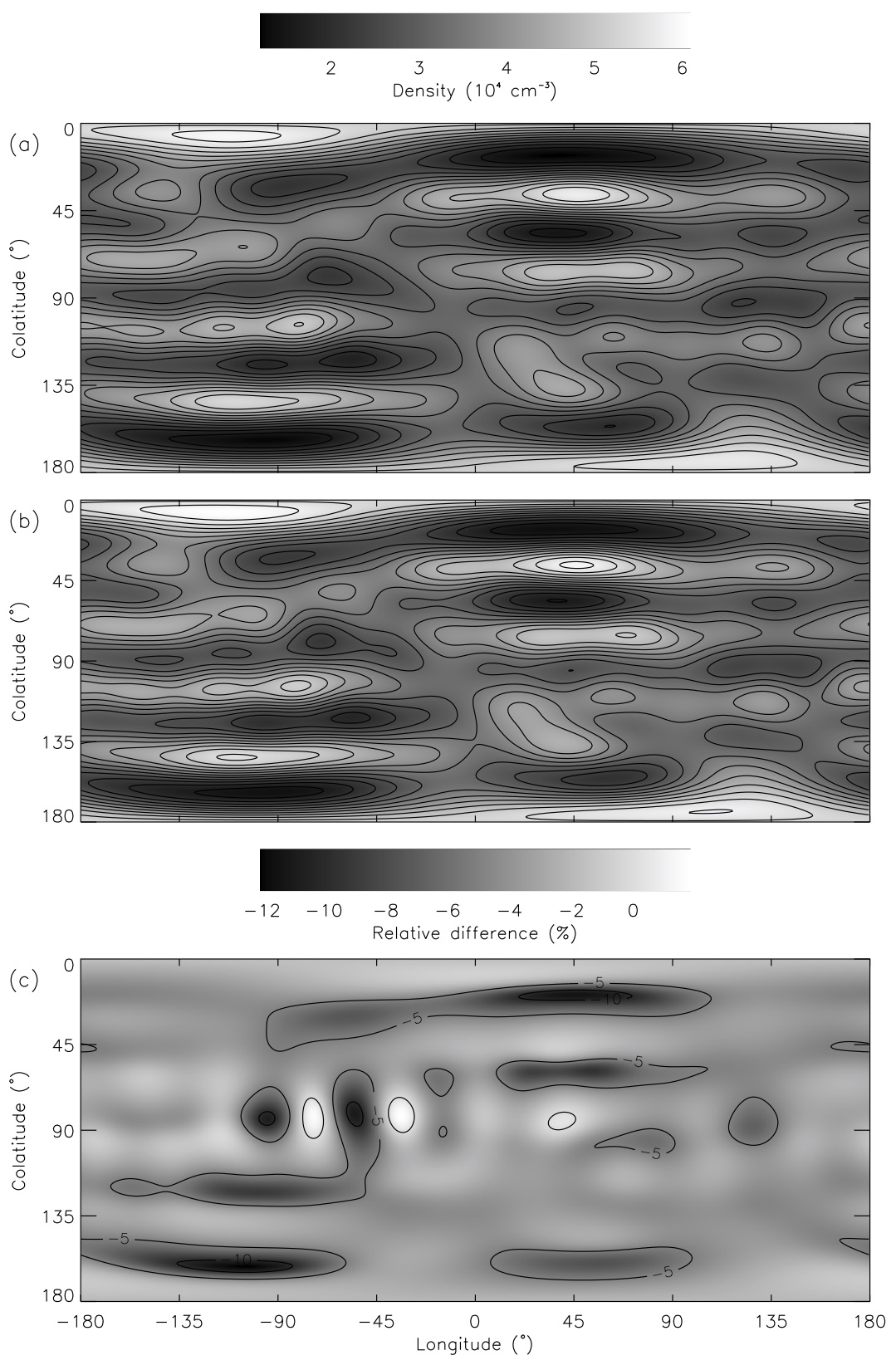}
\end{center}
\caption{(a) The density distribution created using spherical harmonics of order $L=11$ with weighted random coefficients (see text) for a spherical shell at height 5\Rs. (b) The reconstructed density. (c) The percentage difference between target and reconstructed densities. The longitude and colatitudes are Carrington coordinates.}
\label{density0}
\end{figure}

Synthetic observations are made from a known density distribution. For this example, a spherical distribution of density at height 5\Rs\ is created using equation \ref{eq1}, with $L=11$ ($n_{sph}=144$). The $c_i$ are created from a set of random numbers in the range $-1$ to 1, divided by weight $l+m+1$, so that higher-order components are reduced in amplitude. The distribution is then scaled between a minimum at a typical value for electron density in a coronal hole \citep{doyle1999}, and a maximum within a streamer \citep{gibson2003}. This distribution is shown in figure \ref{density0}a. This will be the target density distribution against which the method is tested. The distribution is simple in the sense that it is based directly on spherical harmonics - it is not similar to a true coronal density distribution, yet it serves as an initial test of the method.

\begin{figure}[h]
\begin{center}
\includegraphics[width=8.5cm]{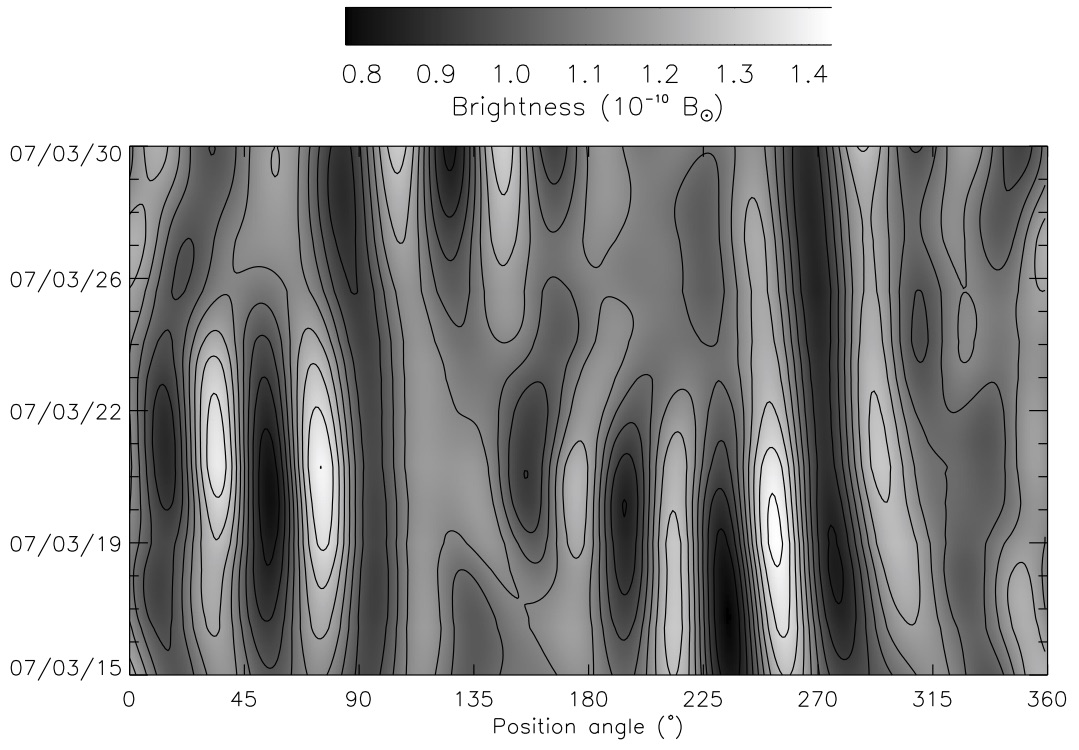}
\end{center}
\caption{$B_k$ values created from the line-of-sight integration of the density distribution of figure \ref{density0}a. The brightness is given for an `observational' height of 5\Rs, giving a synoptic-type map as a function of position angle and time.}
\label{b0}
\end{figure}

Synthetic observations are made by specifying an uniform vector of pixels describing a circle centered on the solar disk as observed from the perspective of LASCO C2 during 2007/03/15-30. One observation per hour is synthesized throughout this period, for 360 pixels distributed at each degree around the circle (or position angle, measured counter clockwise from north). Thus $n_{obs}=\sim1.2 \times 10^5$ pixels are defined. A line of sight (LOS) is created for each pixel, with 200 points along each LOS extending to $\pm 10$\Rs\ from the point of closest approach to the Sun. Appropriate diverging LOS are used (extending in a narrow cone from the position of the coronagraph through the corona). Spherical Carrington coordinates are calculated for each point, and the density set by equation \ref{eq1} and the random coefficients. For this test case, $f(r)$ is not set according to equation \ref{eq2}, since we can directly use the radial description of density decrease with height in a coronal hole given by \citet{doyle1999} to fix the minimum density at each height. Similarly, the formulation of \citet{gibson2003} can be used to set the maximum density at each height. The $g_i$ are then calculated, and the resulting emission summed along each line of sight. The `observed' K-coronal brightness $b$, as a function of position angle and time, is shown in figure \ref{b0}. 

\begin{figure}[h]
\begin{center}
\includegraphics[width=8.5cm]{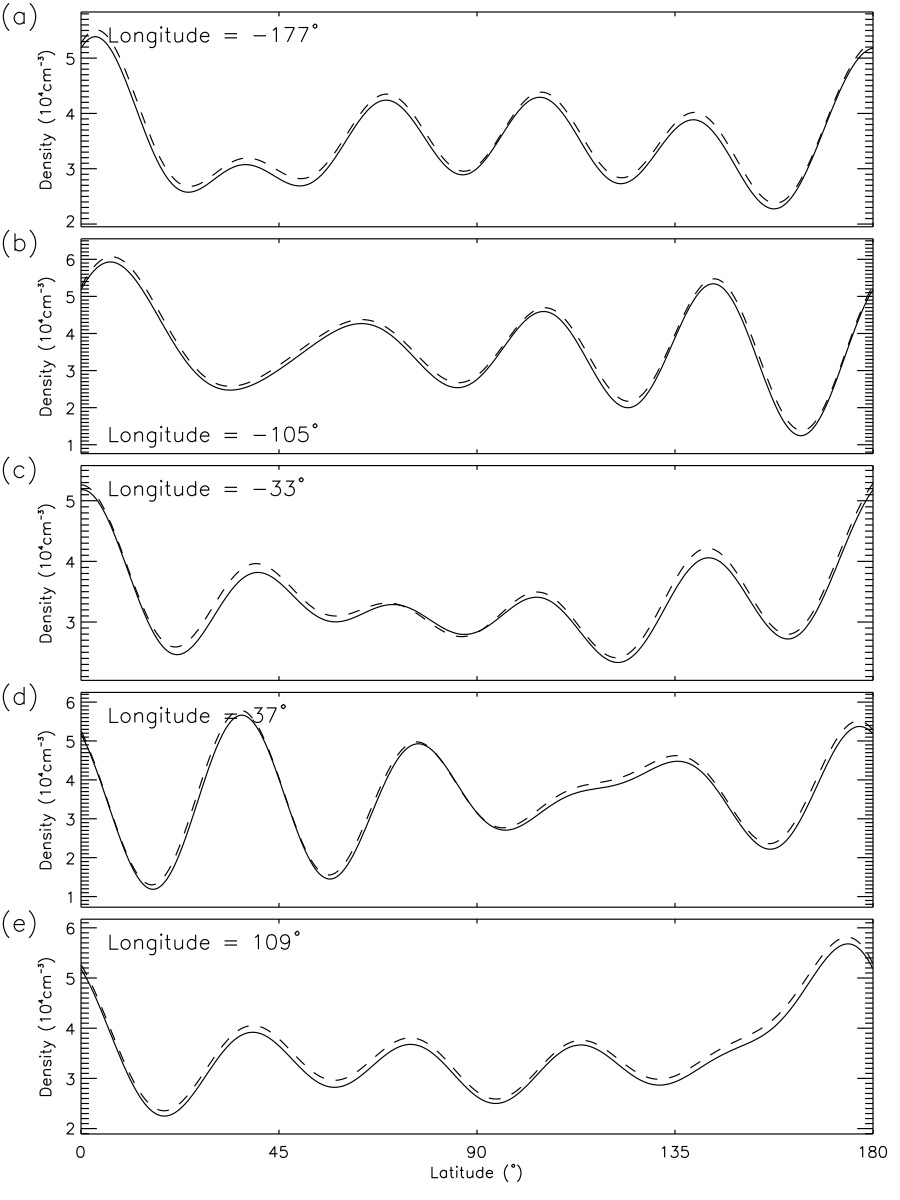}
\end{center}
\caption{Slices of the target density (solid line) and reconstructed density (dashed) as a function of latitude, for various longitudes at a height of 5\Rs.}
\label{dens0slice}
\end{figure}

\begin{figure}[h]
\begin{center}
\includegraphics[width=8.5cm]{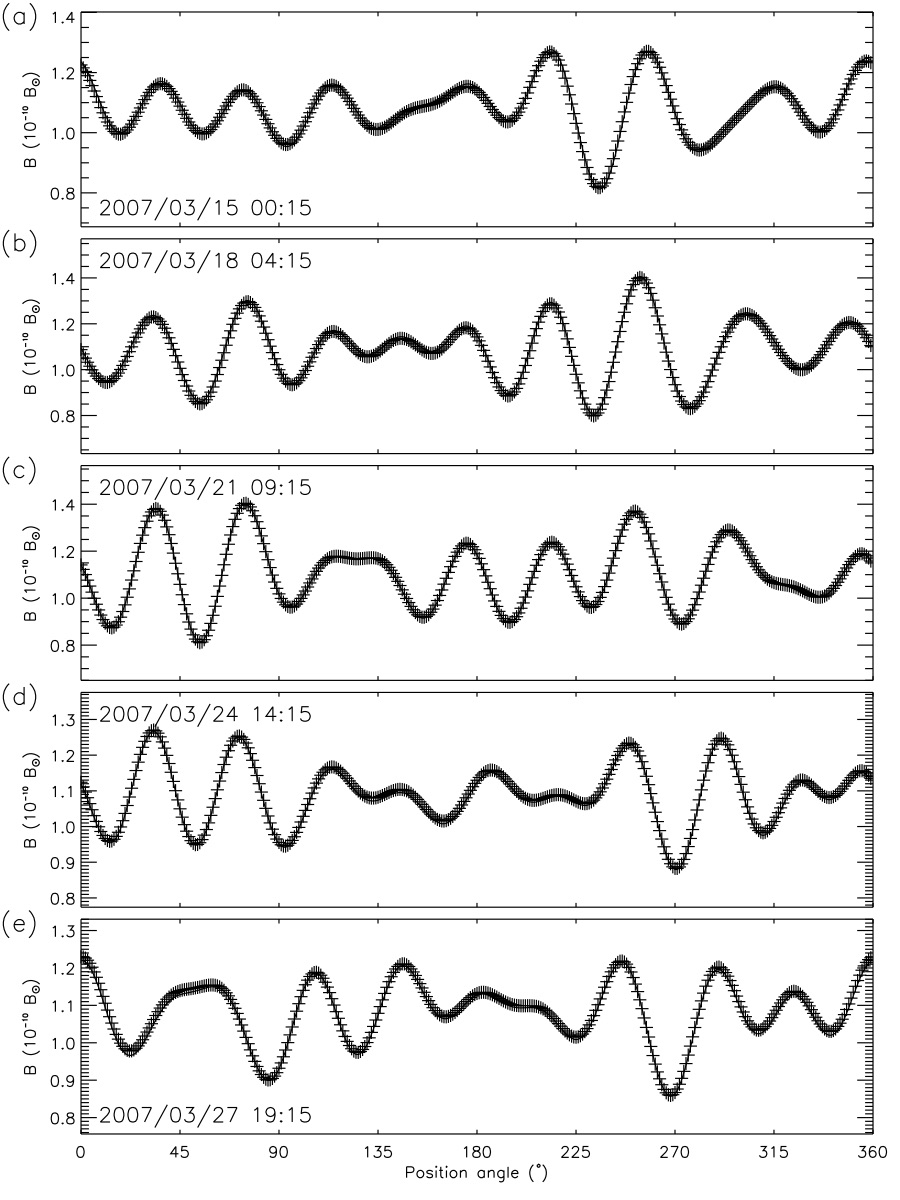}
\end{center}
\caption{Slices of the `observed' (crosses) and reconstructed (line) $B_k$ as a function of position angle, for various `dates' during the test period. They are almost identical.}
\label{b0slice}
\end{figure}

An important choice in reconstructing the density is the choice of $L$, or the maximum number of orders. For the sake of this first simple test, this is set at $L=11$, to match the order of the input distribution. Solving equation \ref{eq8} takes a few seconds on a 2.8GHz Intel Core i7 desktop computer with 16Gb memory. The reconstructed density map is shown in figure \ref{density0}b. The percentage difference between target ($\rho_t$) and reconstructed ($\rho_r$) density is shown in figure \ref{density0}c. The mean absolute percentage deviation is 3.8\%, whilst the distribution correlation $C$ over the sphere, given by

\begin{equation}
C= \frac{\sum (\rho_r-\tilde{\rho}_r)(\rho_t-\tilde{\rho}_t)}{[(\sum (\rho_r-\tilde{\rho}_r)^2)(\sum (\rho_t-\tilde{\rho}_t)^2)]^{0.5}},
\end{equation}
\noindent
is 99.8\% (the $\tilde{\rho}$ are means). Figure \ref{dens0slice} compares latitudinal slices of the observed and reconstructed density for several different longitudes. The residual, or the difference between the reconstructed and observed brightness, is close to zero as shown in figure \ref{b0slice}, which directly compares slices of the observed and reconstructed brightness as a function of position angle for several different times over the `observation' period. The mean absolute fractional deviation of the observed and reconstructed brightness is 0.5\%. 

The algorithm is close to giving a perfect reconstruction for this simple test case. This is perhaps not surprising given that the test density is based directly on spherical harmonics, and that the information on the number of orders ($L=11$) has been used for the solution. Note that the original density distribution used to create the synthetic observations has a density decrease with height based on the formulation of \citet{doyle1999} and \citet{gibson2003}. This gives a decrease with height which is proportional to the relative density of each point, but which does not follow the spherically uniform decrease of equation \ref{eq2}. For the reconstruction, the true decrease is assumed unknown, and equation \ref{eq2} is used. It is obvious from the success of the reconstruction that this leads to only a minor error.

The Appendix describes an alternative method for finding the coefficients $\matr{c}$, based on the properties of the spherical harmonics and iteration. The alternative method performs well in the case where the target density is directly based on spherical harmonics. In general, and for the rest of this work, it is not used since its performance degrades (in both accuracy and efficiency) in comparison to the least-squares method on more complicated density distributions. It is included in the Appendix since it is an interesting approach and may prove useful in other contexts.

\section{A more realistic test}
\label{complex}
\begin{figure}[h]
\begin{center}
\includegraphics[width=8.5cm]{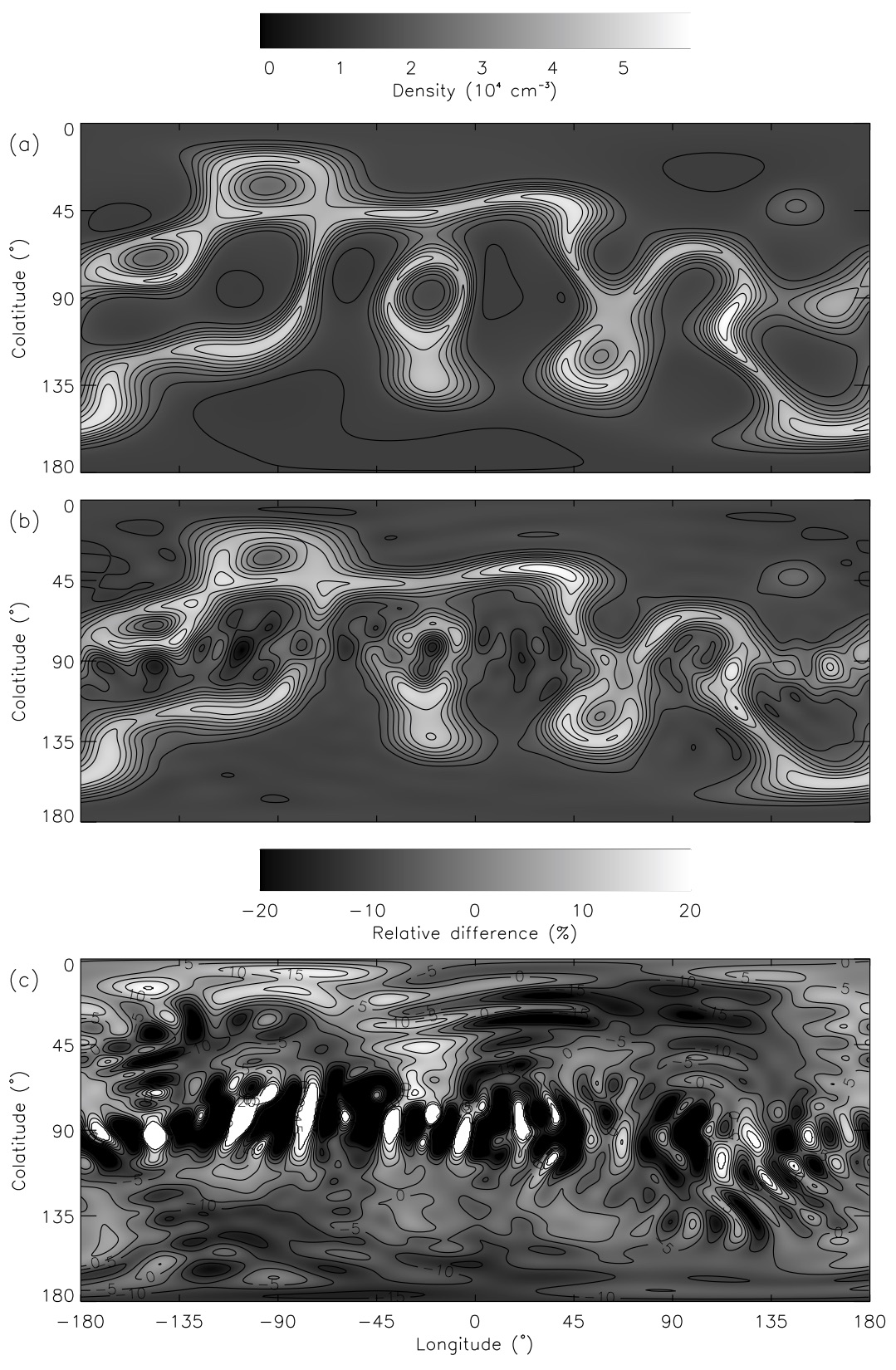}
\end{center}
\caption{As figure \ref{density0}, but for the complicated, narrowly-peaked density distribution of equation \ref{eq14}.}
\label{density1}
\end{figure}

\begin{figure}[h]
\begin{center}
\includegraphics[width=8.5cm]{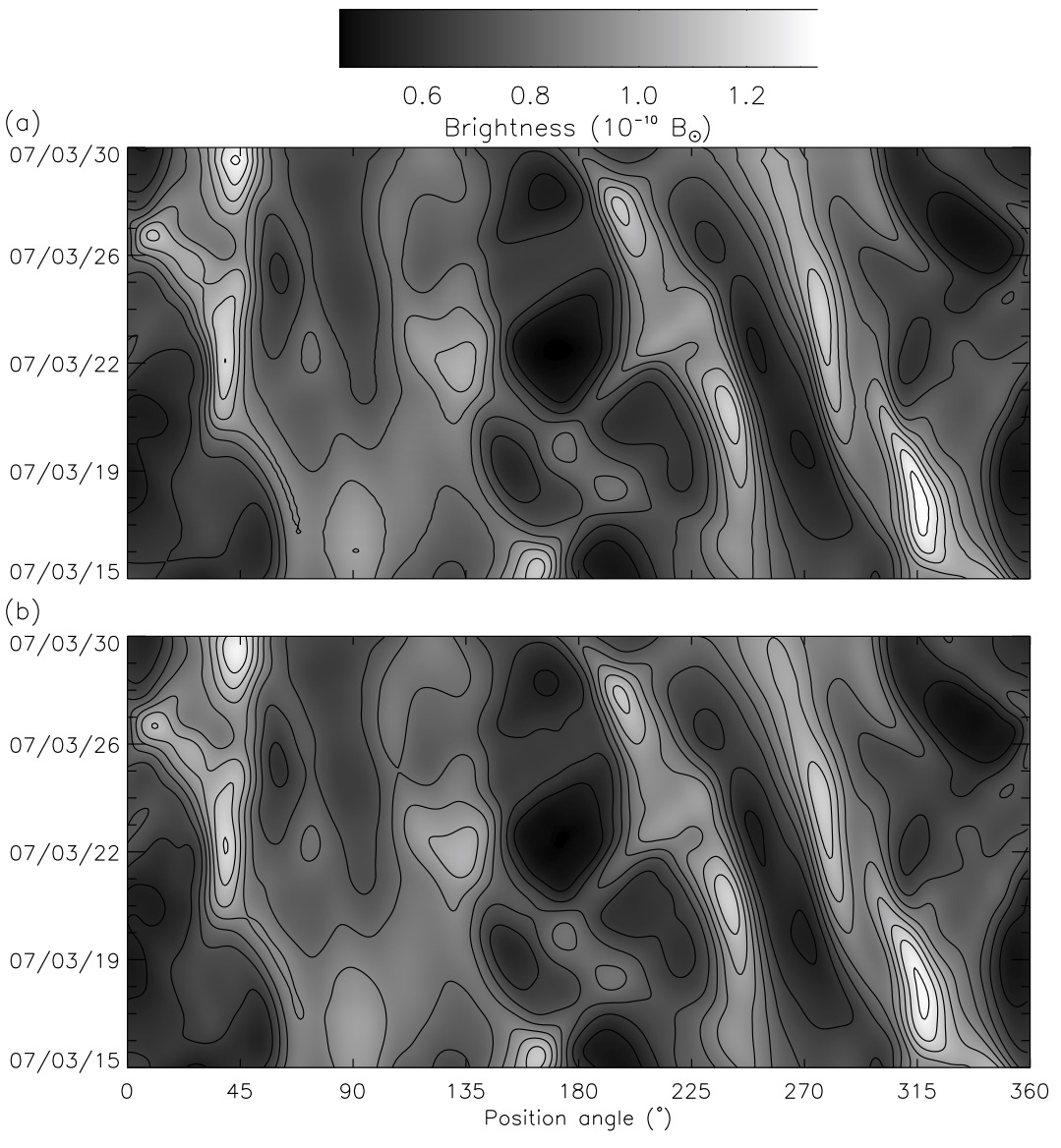}
\end{center}
\caption{(a) $B_k$ values created from the line-of-sight integration of the density distribution of figure \ref{density1}a. The brightness is given for an `observational' height of 5\Rs, giving a synoptic-type map as a function of position angle and time. (b) The model brightness as created from the reconstructed density of figure \ref{density1}b.}
\label{b1}
\end{figure}

In this section, a complicated, narrowly-peaked, density distribution is used to test the reconstruction method. In contrast to the previous simple test, the density distribution is not based directly on a spherical harmonic basis, and therefore the distribution cannot be exactly fitted by a limited order of spherical harmonics, and the number of orders required in the reconstruction cannot be determined beforehand. This distribution is 
\begin{equation}
\label{eq14}
\rho (\phi,\theta)= (\rho_1(\phi,\theta) + 1) \left[ \exp \left( -\frac{\rho_2(\phi,\theta)^2}{\omega} \right)+0.2 \right],
\end{equation}
where $\rho_1$, $\rho_2$ are summed spherical harmonic series with weighted random coefficients (as in the simple case of the preceding section), with $L=11$ and $M=9$, and with $\rho_1$ scaled between 0 and 1. The exponential term forms ridges centered on where the $\rho_2$ function passes through zero, and these ridges can be made narrow by setting $\omega$ to a small value. The $\rho_1$ term introduces variability to the value of both the ridges and the background. This initial density distribution is scaled to appropriate coronal values of density in a similar way to the simple case above. The resulting density distribution is shown in figure \ref{density1}a. Through the exponential function, this distribution has extended, narrow and intricate structures, and is more similar to the expected form of the true coronal density distribution, being distributed along narrow sheets along polarity inversion regions and pseudostreamers \citep[e.g.][]{morgan2010structure}. The brightness resulting from LOS integration of the density is shown in figure \ref{b1}a, again for an `observation' period of half a solar rotation towards the end of March 2007, from the perspective of LASCO C2. 

A high order spherical harmonic basis is required to reconstruct the target density, and for this test we set $L=25$ ($n=676$). The calculation of the LOS integrations of the $A_i$ takes around five minutes on the desktop computer, and the least-squares estimation takes also around five minutes - the calculation of the covariance matrix $\matr{A^\intercal A}$ accounts for most of this time. The reconstructed density has a mean absolute fractional deviation of 13.7\% from the target, with a structural correlation of 94.0\%. The comparison is shown in figure \ref{density1}. The reconstructed brightness, shown in figure \ref{b1}b is almost identical to the `observed', with a mean absolute fractional deviation of 1.2\%.

\begin{figure}[h]
\begin{center}
\includegraphics[width=8.5cm]{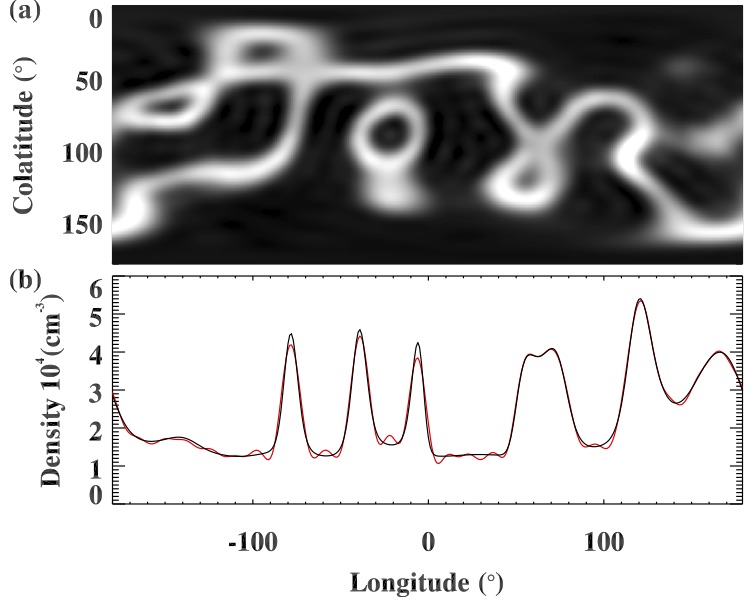}
\end{center}
\caption{(a) The density arising from a direct (non-tomographical) calculation of harmonic coefficients (see text). (b) A slice along the equator comparing true density (black) and spherical harmonic density (red). }
\label{gibbs}
\end{figure}

Despite the decent structural correlation in density distribution, and the almost identical match between model and observed brightness, the reconstruction suffers from high-frequency longitudinal oscillations, leading to large inaccuracy near the equator and regions of low density (including a small negative region). These oscillations are caused by large spherical harmonic coefficient values at high frequencies as the data is overfitted. Figure \ref{gibbs} shows the optimal density that can be achieved using a $25^{th}$ order spherical harmonic basis. The coefficients are calculated directly from integrating the product of each spherical harmonic basis with the true input density over the spherical shell by
\begin{equation}
\label{eqint}
c_i = \int_\phi \int_\theta \rho(\theta,\phi) S_i(\theta,\phi)  \sin\phi \; d_\theta d_\phi .
\end{equation}
Steep jumps in density cause high-frequency oscillations (Gibbs oscillations), which can be seen in figure \ref{gibbs}b. These are minor compared to the large-amplitude errors in the least-squares tomographical reconstruction.

The tendency of the reconstruction to contain negative densities near high-density regions is a problem which plagues coronal tomography. This test shows that it is a problem which arises not solely due to rapid temporal changes in the streamer belt or due to contamination by coronal mass ejections (this test data has zero noise and no temporal changes). It is a problem intrinsic to the observations - of convolution of linear lines of sight through an extended spherical structure, and related to missing information at heights below the height of interest $r_0$ for any single observation. Even for tomography at heights below 5\Rs, this problem is unavoidable at the limit of the instrument field of view. The problem of extreme oscillations in reconstructed density is worse near the equator: for a given observation, the LOS integrations at the equator pass through only a limited range of longitude and through only a very small range of latitude. At the poles, the LOS observations pass through the whole polar corona, near to the axis of rotation, giving a more stable reconstruction. 

The results of this section show that some form of regularization is required to impose smoothness on the reconstruction and to avoid negative densities.

\section{Regularisation of the higher-order harmonics}
\label{regular}

Other coronal tomography methods impose a condition on the spatial smoothness of the reconstruction \citep[e.g.][]{frazin2000} to avoid unphysical high-frequency components. A similar and necessary extension of the spherical harmonic approach is given here. It is desirable to increase the highest order of the spherical harmonics in order to reconstruct the density structure at the finest possible resolution, yet this leads to greater instability of the highest orders. Coronal tomography methods achieve stability by imposing a weighted penalty term for lack of spatial smoothness in the reconstructed density - thus the optimal reconstruction is given by a compromise between the best fit to the data and the spatial smoothness of the reconstruction (regularization). 

The noise $\sigma$ at each position angle and time bin is estimated from the original pre-binned data by isolating the highest-frequency spatial and temporal component. To achieve this, a datacube is created of dimensions position angle, height, and time. The height range is a narrow strip ($\pm 0.2$\Rs) centered on the height of interest. The datacube is convolved with a narrow Gaussian kernel over position angle and time, and this smoothed data subtracted from the original. This leaves the high-frequency residual containing noise, rapid temporal changes, and some residual from very sharp gradients. The narrow height range serves to increase the number of pixels at each point, giving an improved estimate of noise. 

Defining $\matr{A_\sigma}=[\sigma]^{-1}.\matr{A}$ and $\matr{b_\sigma}=[\sigma]^{-1}.\matr{b}$, a regularized solution weighted by the noise reciprocal is given by
\begin{equation}
\label{eq11}
\matr{c=(A_\sigma^\intercal A_\sigma + \lambda w)^{-1}A_\sigma^\intercal b_\sigma},
\end{equation}
where $\lambda$ is a regularisation factor that sets the balance between fitting the data and imposing \emph{a priori} constraints on the solution. $\matr w$ is a square matrix, with diagonal elements $i=0,1,...,n_{sph}-1$ given by
\begin{equation}
\label{eq12}
\matr w_i = \frac{l_i + |m_i|}{\sum_{i}^{n_{sph}-1} l_i + |m_i|},
\end{equation}
and non-diagonal elements are zero (the $l$ and $m$ are the spherical harmonic longitudinal and latitudinal order). $\matr w$ takes the place of the more commonly-used identity matrix so that the regularisation has a larger direct impact on higher frequency harmonics.

In previous work on regularization in coronal tomography, the commonly-used positivity constraint on the density selects values of $\lambda$ where density is everywhere zero or positive. From our own tests on this approach, this gives an overly-smooth solution - that is, for all small values of $\lambda$ the positivity constraint is not satisfied, and only at large values does the density become everywhere positive. A different approach is taken here. Our fitting routine finds an optimal solution using two parameters. One is $\lambda$ (the smoothing parameter), and the other is a minimum density threshold $\rho^\prime$. The main steps in this approach are:
\begin{enumerate}
\item Values $\lambda_k$, with index $k=0,1,...,n_k-1$ are set by a logarithmic increment between the minimum entry of the diagonal of the co-variance matrix $\matr{A_\sigma^\intercal A_\sigma}$ divided by 10, and the maximum entry multiplied by 2. Typically we set $n_k=25$.
\item A minimum density is estimated from the observed brightness values through a spherically-symmetric inversion of the $2^{nd}$ percentile minimum of brightness. Values of $\rho_{j}^\prime$, with index $j=0,1,...,n_j-1$  are set between the minimum density divided by 5, and the minimum density multiplied by 2. Typically we set $n_j=20$.
\item For each value of $\lambda_k$, an initial solution is given by equation \ref{eq11}. This solution gives an initial density distribution on a longitude-latitude map at the coronal height of interest (e.g. 5\Rs). 
\item For each value of $\rho^\prime_j$ the initial reconstruction solution at the current $\lambda_k$ is thresholded to a minimum value of $\rho^\prime_j$. A new set of spherical harmonic coefficients are calculated directly from this thresholded density map via equation \ref{eqint}. These adjusted coefficients $\matr{c}_{k,j}$ are used to give a measure of goodness-of-fit to data for the current value of $\lambda$ and $\rho^\prime$ by:
\begin{equation}
\label{eq13}
\chi_{k,j} =  \frac{1}{n_{obs}} \sum \frac{\sqrt{\left(\matr{b_\sigma-A_\sigma c}_{k,j}\right)^2}}{\matr{\sigma}}.
\end{equation}
\end{enumerate}

\begin{figure}[h]
\begin{center}
\includegraphics[width=8.5cm]{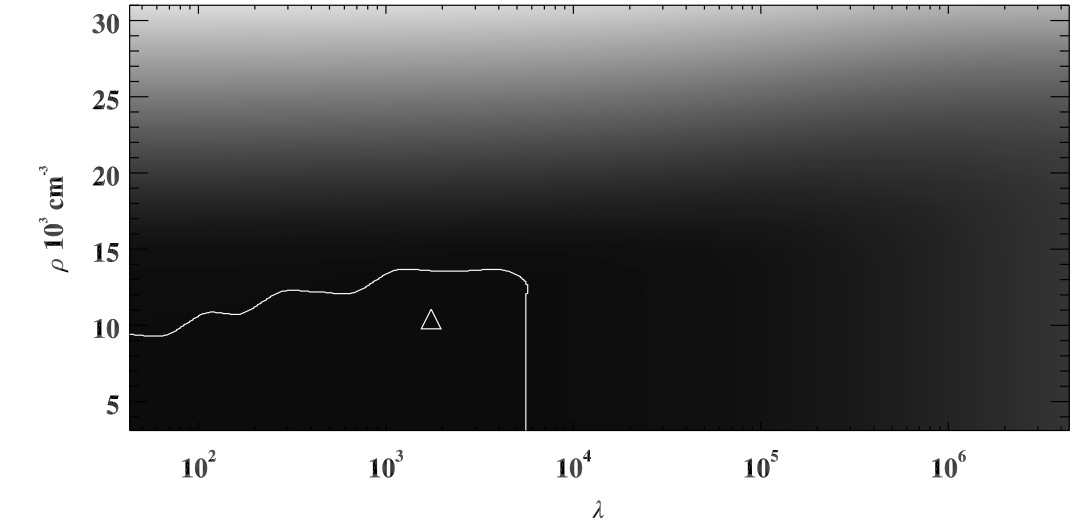}
\end{center}
\caption{The goodness-of-fit to data $\chi_{k,j}$, as defined by equation \ref{eq13} as a function of the regularization parameter $\lambda$ and minimum density threshold $\rho^\prime$. The white contour shows the 15\%\ minimum percentile. The triangle symbol shows the optimal point as described in the text.}
\label{regularfig}
\end{figure}

Thus a 2D array $\chi_{k,j}$ is gained that maps the goodness of fit as a function of $\lambda$ and $\rho^\prime$. The final task is to define an optimal point within this array. Figure \ref{regularfig} shows $\chi_{k,j}$ for the complicated density distribution, calculated over a grid of 25 $\lambda$ and 20 $\rho^\prime$ points. As expected, $\chi$ increases with increasing $\lambda$ - a smoother density reconstruction gives a poorer fit to data. $\chi$ also increases with increasing $\rho^\prime$, since the reconstructed density is thresholded to a higher minimum value, taking it further from the initial least-squares solution. There is a broad region within this array that contains the lowest values of $\chi$ and has very low gradients of $\chi$ (i.e. low variability): $\chi$ increases only slowly in this region as a function of both $\lambda$ and $\rho^\prime$. This region is identified by the 15\%\ percentile minimum value of $\chi$, shown by the white contour. Through tests using several different density distributions, addition of various noise levels (and datagaps), and tests on real data, we define the optimum point within this region as halfway between the region centroid and the point on the region boundary furthest from the origin, shown as the triangle symbol. This point defines our final solution for density. The solution, for this example, has a minimum density threshold of $\rho^\prime=10.4 \times 10^3$\vcm, and $\lambda = 1.74 \times 10^3$ (for interpolated grid position $k=7.65$ and $k=2.98$). The true minimum density of the synthetic density distribution is $1.19 \times 10^3$\vcm. 

\begin{figure}[h]
\begin{center}
\includegraphics[width=8.5cm]{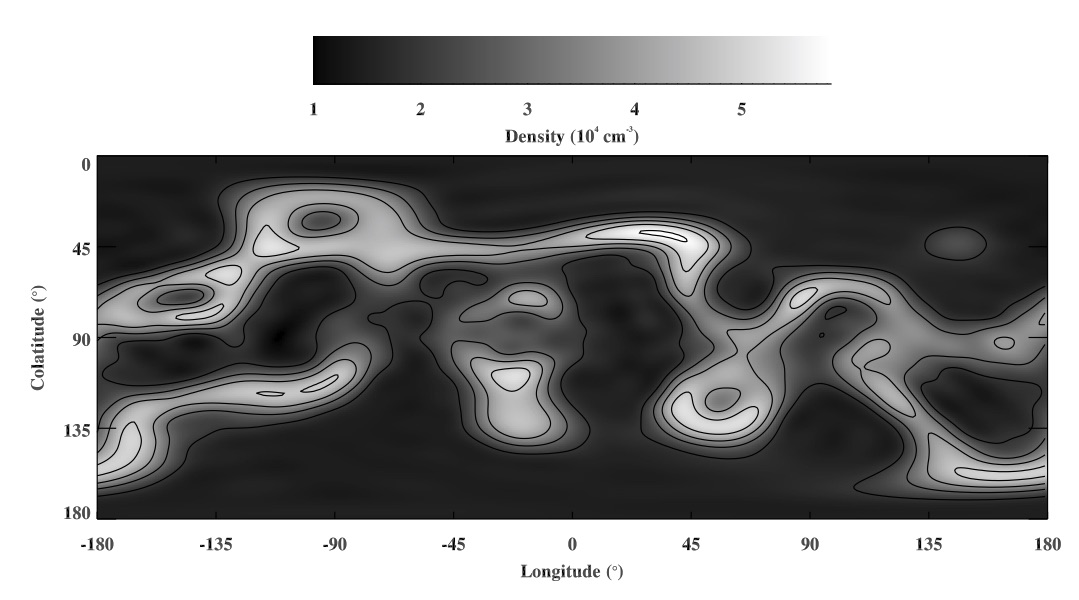}
\end{center}
\caption{The reconstructed density as gained from the regularized fitting method.}
\label{densweights}
\end{figure}

Application of this fitting routine results in a considerable improvement in reconstructed density, as shown in figure \ref{densweights}. The high-frequency oscillations near the equator are greatly reduced. The density has a mean absolute fractional deviation of 12.3\% from the target, with a structural correlation of 95\%. The brightness values are fitted with a mean absolute deviation of 1.1\%. As inherent to the fitting method, there are no regions of negative density. The fitting routine adds around 5 minutes to the computational time: the $\matr{A^\intercal A}$ covariance matrix is pre-computed, and calculations of modeled brightness and density for equations \ref{eq13} and \ref{eq14} are efficient due to the spherical harmonic basis. Note that for this test case, there is no noise, so an arbitrary constant value of noise is set for each data point (i.e. no weighting in equation \ref{eq11}).

\section{Missing data, noise and rapid temporal changes}
\label{noise}

\begin{figure}[h]
\begin{center}
\includegraphics[width=8.5cm]{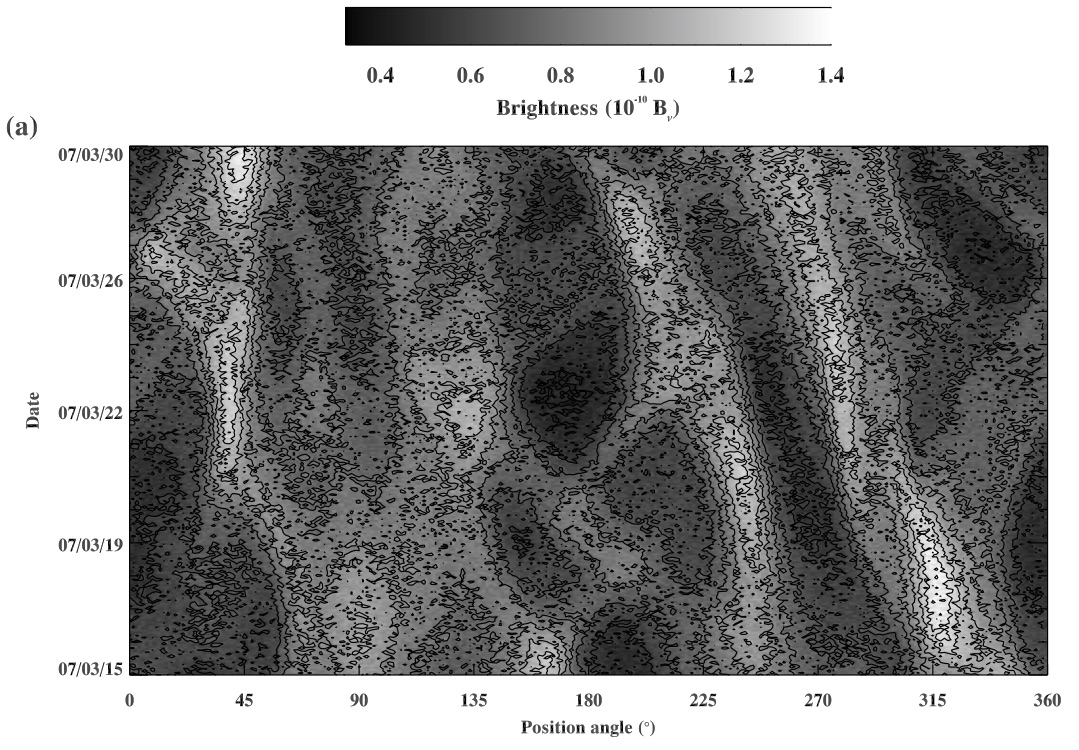}
\includegraphics[width=8.5cm]{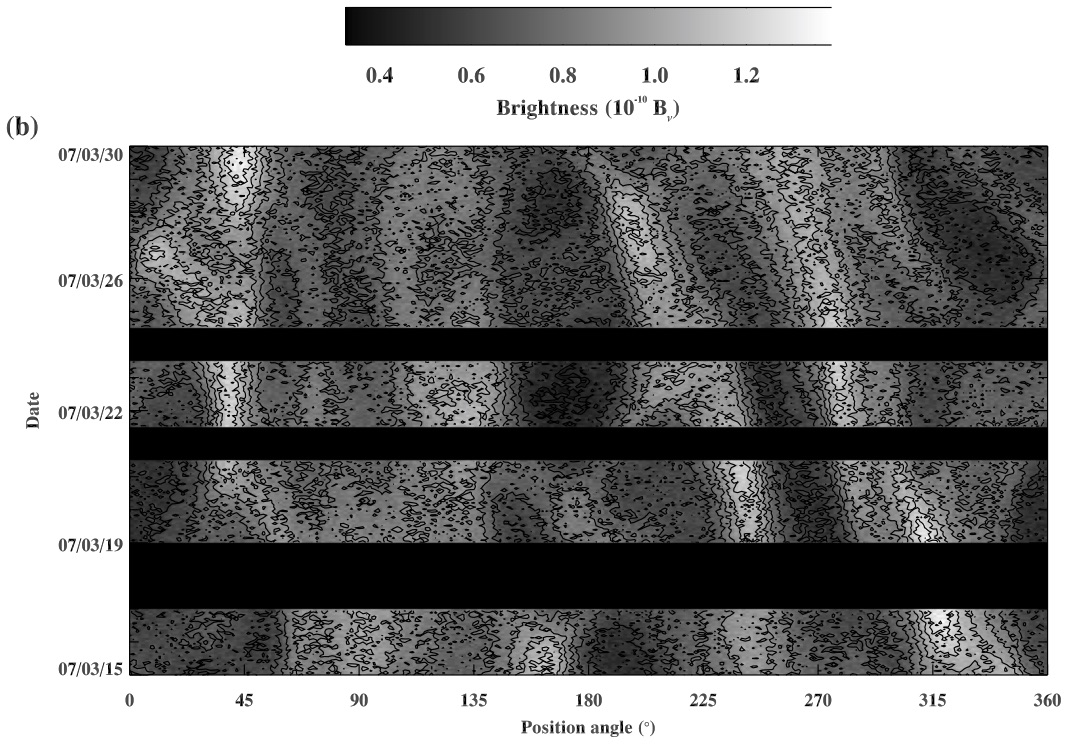}
\end{center}
\caption{(a) The synthetic brightness data degraded by 5\%\ normally-distributed random noise. (b) A set of synthetic observations with three periods of missing data (rectangular black blocks) centered on 2007/03/18, 22 and 25. The first period lasts for two days, the two other periods for a day each. Noise with amplitude 5\%\ of the mean signal is also present in this data.}
\label{bnoise}
\end{figure}

\begin{figure}[h]
\begin{center}
\includegraphics[width=8.5cm]{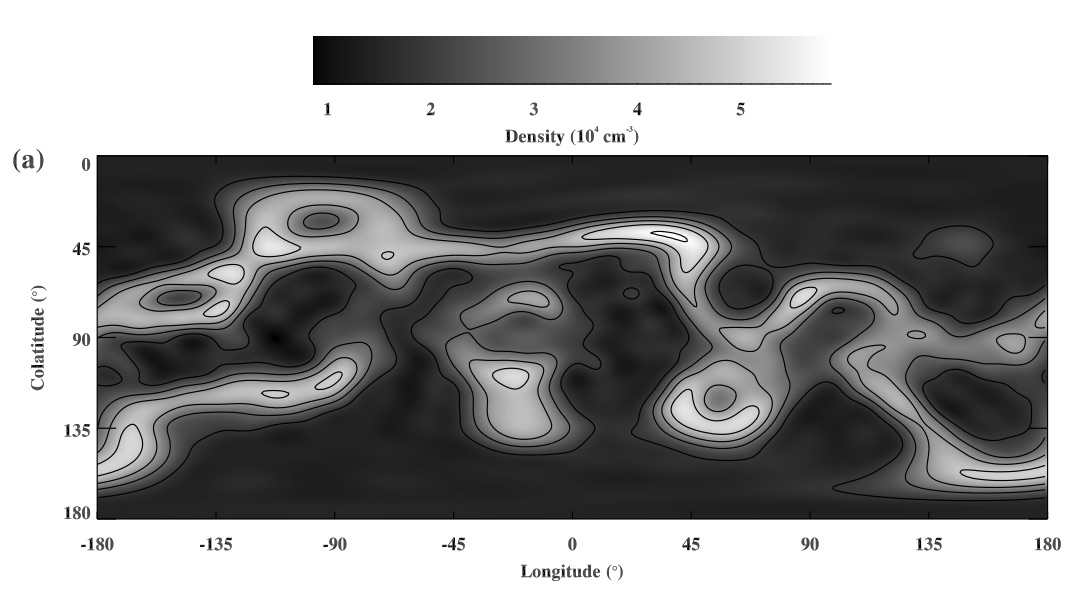}
\includegraphics[width=8.5cm]{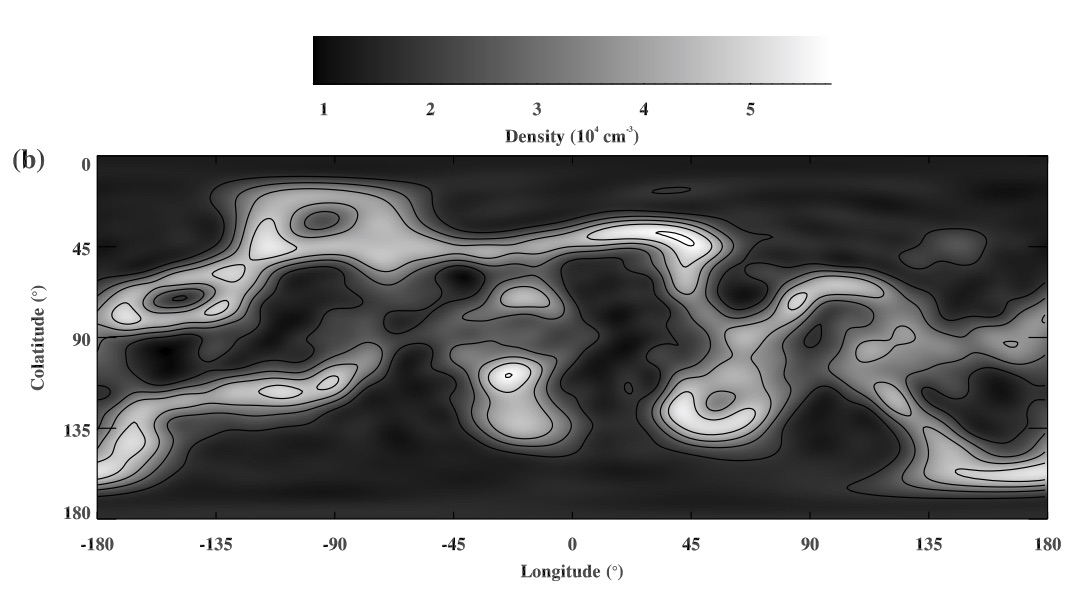}
\end{center}
\caption{(a) The reconstructed density for the input data degraded by noise. (b) As (a), but for the noisy input data including datagaps.}
\label{densnoise}
\end{figure}

Figure \ref{bnoise}a shows the brightness test data degraded through the addition of random normally-distributed noise at 5\%\ of the mean signal level. Regularized tomography applied to this noisy dataset gives the density of figure \ref{densnoise}a. The reconstructed density has a mean absolute fractional deviation of 12.1\% from the target, with a structural correlation of 95\%. The brightness values are fitted with a mean absolute deviation of 4.3\%. The solution has a minimum density of $\rho^\prime=9.96 \times 10^3$\vcm, and $\lambda = 1.75 \times 10^3$.

The largest reconstruction errors are near the equator, where high-density regions are underestimated, and low-density regions overestimated - that is, the reconstruction gives density which is too smooth over longitude compared to the sharply-defined structures and large gradients of the true density. This is an important point to remember when interpreting tomography results applied to real data - the equatorial regions are the most important regions in the context of space weather studies, yet is where the reconstruction errors are greatest.

All coronagraphs suffer from occasional datagaps, with the potential to seriously degrade tomographical reconstructions. Figure \ref{bnoise}b shows a half-solar-rotation set of noisy synthetic observations with 4 missing days of data (around one-third missing) split into 3 gaps of 2 days, 1 day and 1 day. The reconstructed density for this data is shown in figure \ref{densnoise}b. It deviates from the target density by 14.1\%, with a spatial correlation of 94\%. The reconstructed and observed brightness deviate by 4.3\%. The solution has a minimum density of $\rho^\prime=9.9 \times 10^3$\vcm, and $\lambda = 2.19 \times 10^3$. Thus the spherical harmonic basis provides stability in the presence of even quite substantial datagaps. 

The most detrimental noise in coronagraph data is perhaps not a normal distribution, but rather isolated pixels or groups of pixels of spurious high/low values caused by, for example, sporadic bursts of energetic particles which can seriously deteriorate some images, or the passage of bright planets. The weighted fitting can help reduce the impact of these on the results. More importantly, rapid changes in brightness and structure caused by CMEs have a large detrimental effect on reconstruction. \paperi\ introduces several processing steps to reduce these problems. In particular, the dynamic separation technique (DST) reduces the effect of CMEs, and also results in a smoother signal with reduced salt-and-pepper noise. Observations which are seriously degraded (possibly due to bursts of energetic particles), can be identified and discarded, as described in \paperi. Occasionally, telemetry or read errors can lead to missing blocks of data within an image. Discarding bad images, or missing data blocks, will result in short datagaps, which seems acceptable for the spherical harmonic method as shown above.

Lastly, coronal structure must change, either slowly, or rapidly, and may reconfigure very rapidly during the passage of large CMEs. Time-dependent coronal tomography (based on regularisation methods) has been successfully applied by \citet{vibert2016}. In principle, the spherical harmonic approach can be extended to include time-dependency, with the coefficients becoming functions of time. Initial experiments with a time-dependent density model shows that this is a very challenging task - particularly if a step-change in density is needed to account for rapid changes. Further development is necessary, reserved for a future publication.

\section{Application to observations}
\label{observations}
\begin{figure}[h]
\begin{center}
\includegraphics[width=8.5cm]{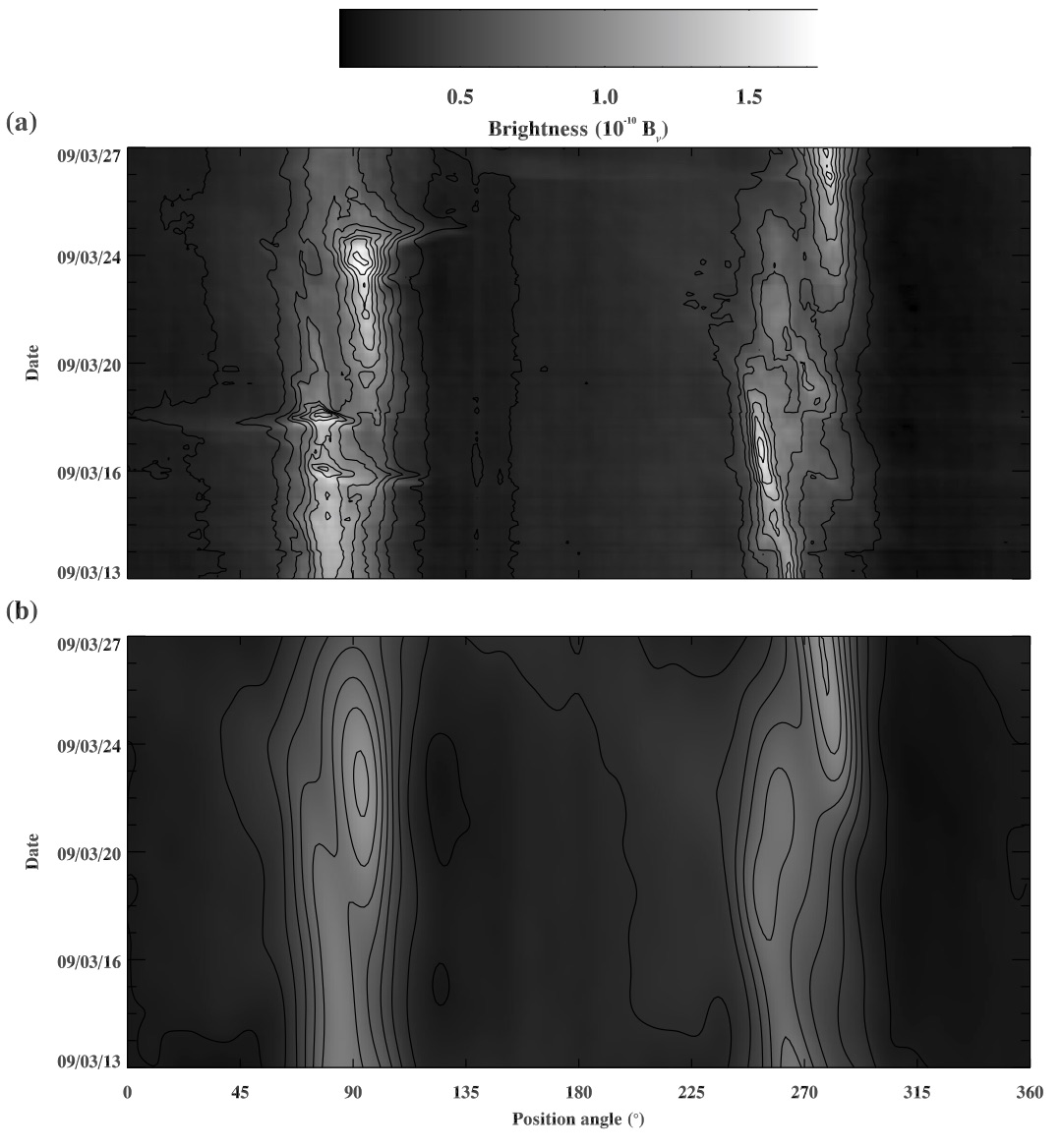}
\includegraphics[width=8.5cm]{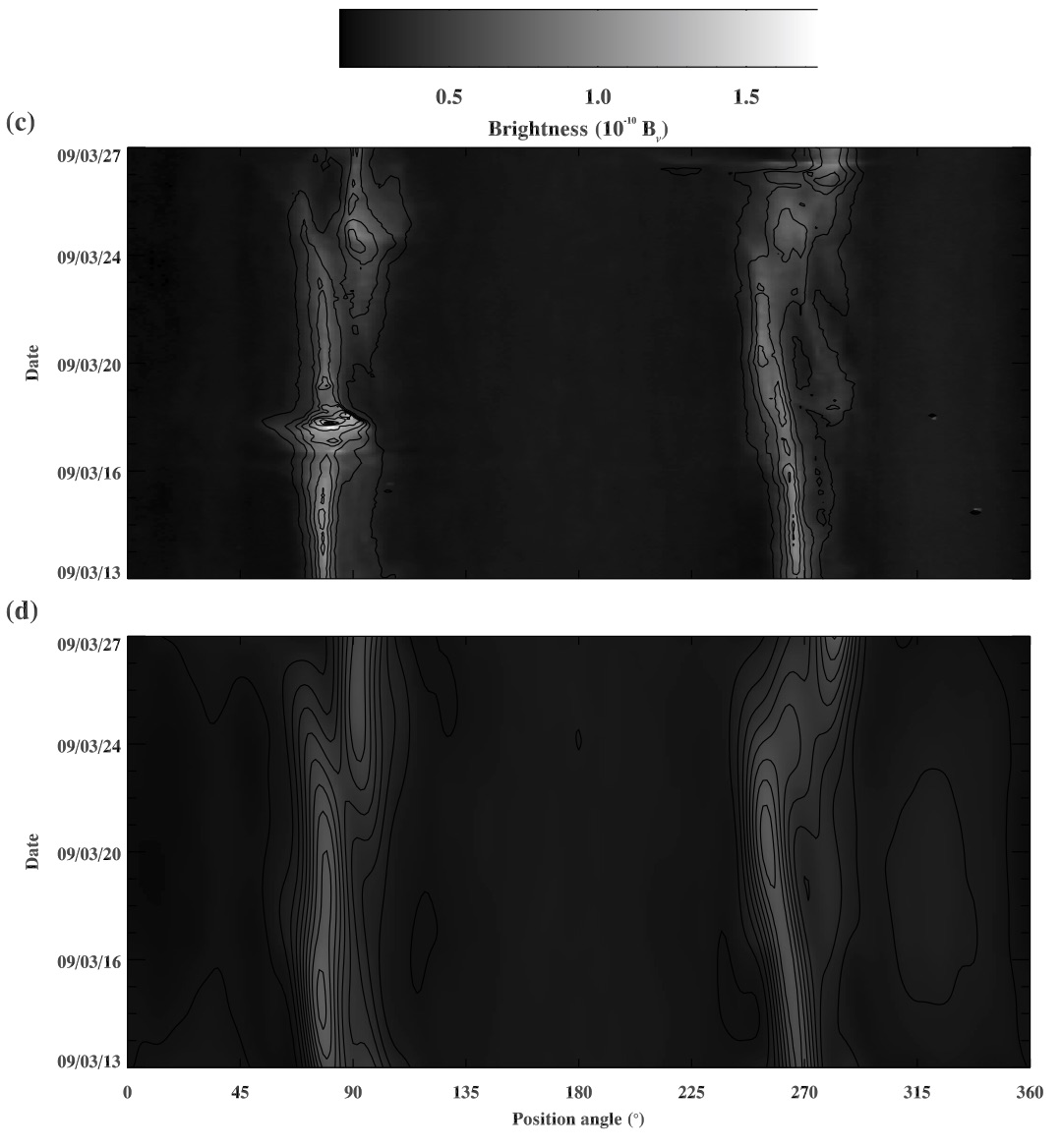}
\end{center}
\caption{(a) The brightness of the corona observed at 5.5\Rs\ by LASCO C2 for a two-week period centered on 2009/03/20. (b) Model brightness gained from reconstructed density for LASCO C2. (c) As (a), but observed by the COR2 A instrument. (d) Model brightness for the COR2 A reconstructed density.}
\label{realdata}
\end{figure}

This section applies the tomography to observations made by the LASCO C2 and the STEREO SECCHI COR2 A coronagraphs for a half-Carrington rotation period centered on 2009/03/20 12:00. At this time, the STEREO A spacecraft is separated by $60^\circ$ from SOHO. The data are processed and calibrated according to the method of \paperi. The height of interest is set at 5.5\Rs, and the data rebinned into a position-angle and time array with 180 position angle bins, 200 time steps. The data array for LASCO C2 is shown in figure \ref{realdata}a, and for COR2 A in figure \ref{realdata}c. The data binning can be set at higher resolution, at the expense of computational time. The binning here allows reconstructions to be made in approximately 10 minutes.

\begin{figure}[h]
\begin{center}
\includegraphics[width=8.5cm]{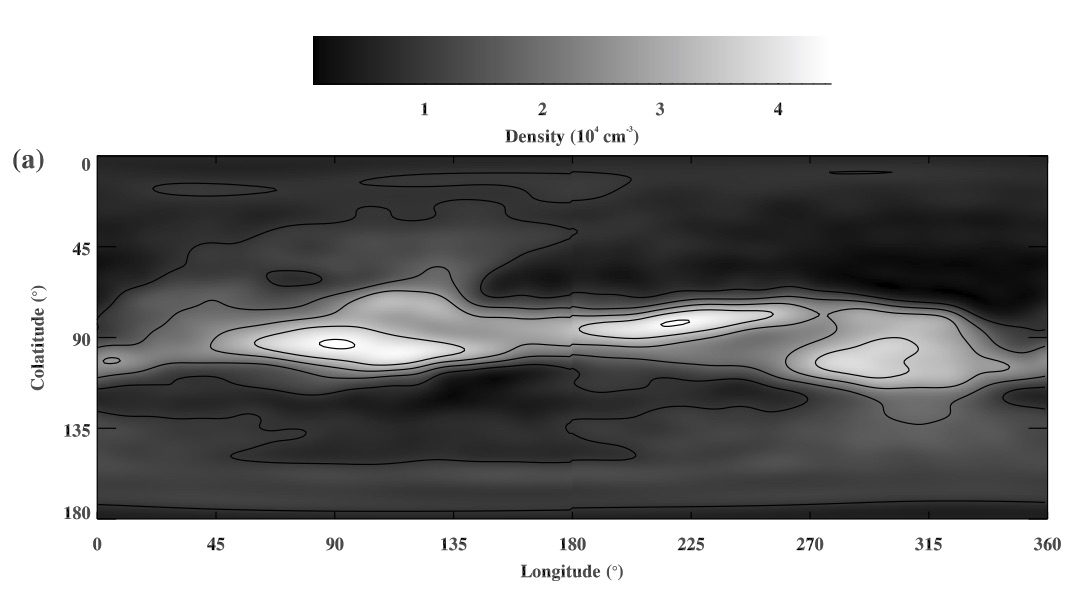}
\includegraphics[width=8.5cm]{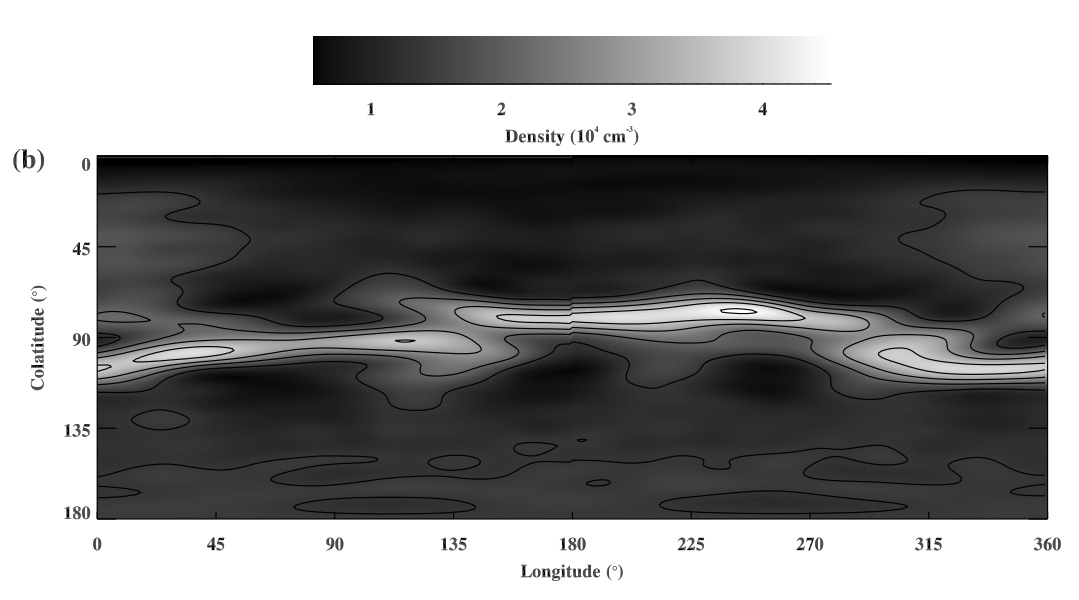}
\end{center}
\caption{(a) Reconstructed density at a height of 5.5\Rs\ gained from the LASCO C2 observations shown in figure \ref{realdata}a. (b) As (a), but for the COR2 A observations shown in figure \ref{realdata}c.}
\label{realdens}
\end{figure}

The choice of period, and height, is to allow convenient comparison with figure 5 of \citet{frazin2010}. The density reconstruction for LASCO C2 is shown in figure \ref{realdens}a, and for COR2 A in figure \ref{realdens}b. The LASCO C2 data is fitted with a mean absolute deviation of 10.6\%, with a smoothing parameter of $\lambda=6.2 \times 10^4$ and minimum density $\rho_{min}=1.4 \times 10^3$\vcm. For COR2 A the values are 7.0\%, $\lambda=5.1 \times 10^4$ and $\rho_{min}=6.5 \times 10^3$\vcm. The mean absolute fractional difference between the two reconstructed densities is 38\%, with a spatial correlation of 81\%. Comparing with figure 5 of \citet{frazin2010}, these density maps are smoother, and have maximum densities at around half the values of \citet{frazin2010}. Currently there is no other empirical verification for density maps such as these. From figure \ref{realdens}, COR2 A seems to give a better reconstruction, in that the streamer belt is narrow, and is fitting the data more closely. Comparison with future \emph{in situ} measurements of the coronal density by the Parker Solar Probe will be invaluable for coronal tomography.

\section{Conclusions and future work}
\label{conclusions}

For heights where the coronal structure can be well-approximated as radial with an uniform density decrease with increasing height (i.e. the extended inner corona), a model of the density based on spherical harmonics leads to a very efficient and stable method for reconstruction. This is demonstrated for a simple and complex model coronal density distribution. The method is robust to large datagaps of several days. Without regularisation, the smoothness of the reconstructed density is dictated by the highest order of the spherical harmonic basis. However, the true coronal density is likely to have steep gradients between regions of low and high density, or very narrow regions of high density, and a high order is required to approximate these. To counteract this problem, we provide a method for regularised solutions where the smoothness of the reconstructed density, and a minimum density threshold, is taken into consideration. 

The application of this method to a large dataset will be presented in the third paper of this series. Other future work involves finding a robust time-dependent extension to the spherical harmonic approach, where the harmonic coefficients can change as a function of time. We also aim to experiment with other approaches similar to spherical harmonics that have proved useful in geophysics, including wavelet-based spherical functions \citep{chambodut2005}. We anticipate these may prove useful for the non-radial corona, in particular for extreme ultraviolet (EUV) observations of the low corona.

Spherical harmonics are a simple yet powerful basis for inversion of coronal density, and should be a consideration for other coronal applications such as EUV diagnostics in the low corona, or 3D reconstructions of the coronal magnetic field with future spectropolarimetric instruments. 

\hskip 20pt
\appendix{}
\centerline{\bf Appendix}
\label{appendixa}
This appendix describes an iterative procedure to find the spherical harmonic coefficients $c_i$. For this procedure, the observed data $b$ and the $A_i$ (see equation \ref{eq6} of section \ref{outline}) are first normalized to achieve numerical stability - both are very small numbers ($b$ and $|A_i|$ on the order of $10^{-10}$ and $10^{-16}$  respectively). $b$ is normalized to a mean of zero and unity standard deviation by
\begin{equation}
\label{eq1app}
b^{\prime}=\frac{b-\tilde{b}}{\sigma_{b}}
\end{equation}
where $\tilde{b}$ is the mean and $\sigma_{b}$ is the standard deviation. The $A_i$ are normalized by the mean of their absolute value (calculated over all orders):
\begin{equation}
\label{eq2app}
A_i^{\prime}=\frac{A_i}{<|A|>} .
\end{equation}

Starting with an initial estimate of coefficients (labelled with a prime, $c_i^\prime$, since they are operating on normalized arrays) all set to zero, the following iterative algorithm, with iteration counter $k$ cycling through equations \ref{eq3app}-\ref{eq5app}, converges towards a solution:
\begin{equation}
\label{eq3app}
b_{mod} = \sum_{i=1}^{n_{sph}} c_{i(k)}^\prime A_{i}^\prime,
\end{equation}
\begin{equation}
\label{eq4app}
c_{i(k+1)}^\prime= c_{i(k)}^\prime + \frac{\lambda}{n_{obs}} \sum_{p=1}^{nobs}A_{i}^\prime \left( b^\prime - b_{mod} \right),
\end{equation}
\begin{equation}
\label{eq5app}
c_{i(k+1)}^\prime= \frac{c_{i(k+1)}^\prime }{\sigma_{b_{mod}}},
\end{equation}
where $\sigma_{b_{mod}}$ is the standard deviation of $b_{mod}$ and $\lambda\ (\ll1)$ is a parameter that controls the rate of convergence. At values too large, the process does not converge. This becomes important as the number of spherical harmonic orders becomes high. $\lambda=\frac{1}{n_{sph}}$ gives good results for the examples in this work (where $n_{sph}$ is the number of spherical harmonics). The iterations continue until $k$ reaches a set value, or the convergence rate drops below a set threshold. Note that equation \ref{eq5app} is not strictly necessary, it is included to greatly increase the rate of convergence.

After convergence is reached, the $c_i^\prime$ are scaled to account for the normalizations of equations \ref{eq1app} and \ref{eq2app}, to give solution $c_i$:
\begin{equation}
\label{eq6app}
c_i= \frac{c_i^\prime\ \sigma_{b}}{\sigma_{b_{mod}}}, 
\end{equation}
where $\sigma_{b}$ and $\sigma_{b_{mod}}$ are the standard deviations of the observed and modelled brightness. Finally, the mean density which should be included in the zeroth-order DC component, $c_0$, is estimated directly from the observed brightness by:
\begin{equation}
\label{eq7app}
c_0= \frac{C}{n_{obs}} \sum_{p=1}^{n_{obs}} \frac{b}{\sum_{j=1}^{n_{los}}g_j f(r_j)}.
\end{equation}
$C$ is a correction factor based on the curtailing of the line of sight to a limited range. Due to the curtailing, the summation in the denominator is too small, leading to an overestimate of the mean density by a few percent. This correction is easily quantified by calculating $\sum_{j=1}^{n_{los}}g_j f(r_j)$ for a single case of a very long line of sight (where the emission essentially drops to zero at large heights), and comparing the same value for the curtailed line of sight. This gives the correction factor $C$ directly.

To fit any function on a sphere to a set of spherical harmonics, the coefficient of a spherical harmonic at a given order can be found by integrating the product of the function and the spherical harmonic over the sphere (see equation \ref{eqint}). In this case, where the spherical harmonics are multiplied by geometrical and other factors and integrated over extended lines of sight, the iterative algorithm of equations \ref{eq3app}-\ref{eq5app} in essence implements a similar approach. For the simple test case of section \ref{simple}, this iterative method gains a more accurate reconstruction than the least-squares approach, with a mean absolute fractional deviation of 2\% between the reconstructed and target densities. For the more complicated cases, it loses accuracy compared to the least-squares approach, and with increasing number of spherical harmonic orders, it becomes considerably less efficient.

\end{document}